%Paper: hep-th/9503145
%From: banks@MAILBOX.SLAC.Stanford.EDU
%Date: Tue, 21 Mar 1995 14:14:38 -0800
%Date (revised): Tue, 28 Mar 1995 12:10:50 -0800

\input harvmac

\Title{RU-95-04}{\vbox{\centerline{Vertex Operators in $2K$ Dimensions} }}
\smallskip
\centerline{\it
T. Banks  \footnote{*}
{\rm  John S. Guggenheim Foundation Fellow, 1994-95;
Varon Visiting Professor at the Weizmann Institute of Science;
Supported in part by the Department of Energy under grant No.
DE-FG05-90ER40559  .} }
\smallskip
\centerline{Department of Physics and Astronomy}
\centerline{Rutgers University}
\centerline{Piscataway, NJ 08855-0849}

\noindent
\bigskip
%\vskip 2cm
\baselineskip 18pt

\vfill
\noindent
A formula is proposed which expresses free fermion fields in $2K$
dimensions in terms of the Cartan currents of the free fermion
current algebra.  This leads, in an obvious manner,
 to a vertex operator construction of nonabelian free fermion current
algebras in arbitary even dimension.  It is conjectured that these ideas
may generalize to a wide class of conformal field theories.
\Date{March 1995}
%\draftmode
\eject

\newsec{Introduction}

The vertex operator construction of two dimensional current algebras,
and the associated procedure of bosonization of fermions\ref\vert{P.Jordan,
E.Wigner, {\it Zeit. Physik.}{\bf 47},(1928),631; E.Lieb,
T.Schultz, D.Mattis, {\it Ann. Phys.}{\bf 16},(1961),407; E.Lieb, D.Mattis,
{\it Phys. Rev.}{\bf 25},(1962),164;
S.Coleman,{\it Phys. Rev.}{\bf D11},(1975), 2088; S.Mandelstam{\it Phys. Rev.}
{\bf D11},(1975),3026;M.B.Halpern,{\it Phys. Rev.}{\bf D12},(1975),1684
{\it Phys. Rev.}{\bf D13},(1976),337; T.Banks,D.Horn,H.Neuberger,
{\it Nucl.\break Phys.}{\bf B}, (1976);  I.Frenkel,
V.Kac {\it Invent. Math. }{\bf 62},(1980),23; G Segal {\it Comm.
Math. Phys.}{\bf 80},(1981),301.}
are powerful tools for obtaining nonperturbative information about
two dimensional field theory.  The purpose of the present note is to
derive an analogous formula in any number of dimensions, in particular
$4$.  We will present precise results only for theories of free
fermions, although in the penultimate section we will also speculate
about current algebras in nontrivial conformal field theories.

The basic feature of the vertex operator construction is that it
enables us to construct charged operators out of the diagonal currents
of the algebra, which are themselves uncharged under the Cartan
subalgebra.  A hint that a similar construction may be possible
in other dimensions is Schwinger's proof that an
abelian spatial current density always carries charge density
in a positive metric quantum field theory.  Formally the Schwinger
charge density integrates to zero, but one might imagine that a
sufficiently nonlocal functional of the current might actually
carry charge.  The challenge is to find such a functional which is
also a local operator.

The real clue to generalizing the bosonization formulae to higher
dimensions comes from a convenient rewriting of the two dimensional
formulae.  Recall that a Weyl fermion can be constructed as
\eqn\twodferm{\psi (z) = e^{-i\phi (z)}}
where the holomorphic current operator $j(z)$ is
\eqn\current{j(z) = i\partial_z \phi (z)}
We rewrite this as
\eqn\twodfermb{\psi (z) = e^{i\int A_{\mu}(w,\bar{w} ) j^{\mu}(w,\bar{w}
) d^2 w}}
where
\eqn\vecpota{A_w = {1\over (w - z)}}
\eqn\vecpotb{A_{\bar{w}} = {1\over (\bar{w} - \bar{z} )}}
Notice that this vector potential has delta function field strength
concentrated at $z$.  It is a singular instanton.

The fundamental reason that bosonization works then is that the
Ward identity for a charged field
\eqn\ward{\partial_{\bar{z}} j(z) \psi (w) = \delta^2 (z - w) \psi (w)}
is just the anomaly equation
\eqn\anomeq{\partial_{\bar{z}} j(z)  = \epsilon^{\mu\nu} F_{\mu\nu}(z)}
for the chiral current in an instanton background.  In order to
generalize
this observation to multiple field correlators, one uses the abelian
nature of the background gauge field.

Another aspect of this two dimensional observation that will be of use
in higher dimensions is the following.  What we have said is that
studying the insertion of a single fermion operator at $z$ is the same
as studying the theory in the background of a singular instanton located
at $z$.  Consider multifermion correlators in such a background.
$U(1)$ charge conservation plus the fact that the fermions are free,
tell us that the only nonvanishing connected correlation function is the
one point function of the conjugate fermion field $\bar{\psi}$.

To obtain the fermion two point function, we consider the {\it partition
function} of the system in the presence of a singular instanton-antiinstanton
pair located at $0$ and $z$.  This is the determinant of the Weyl operator
and requires careful definition.  We define the real part of the log of the
determinant as one half the Schwinger result for a full Dirac fermion
(with the gauge field coupled to the vector current).  The imaginary part is
defined by the anomaly equation. The singular instanton is almost everywhere
a pure gauge, and the variation of the determinant with respect to the gauge
function is $\int i\delta\theta \rho_{top}$, where $\rho_{top}$ is the
topological charge density.  Plugging in the singular gauge function
and topological
charge density of an instanton anti-instanton pair, we determine the
phase of the determinant.  In this calculation, infinite, $z$ independent,
constants in the log of the determinant
are dropped and the normalization of the two point function is
thus determined by hand.  This procedure gives the correct result ${1\over z}$.
Multipoint correlators are obtained from multiple instanton pairs.  They
produce the correct free fermion behavior as a consequence of the bilinearity
of Schwinger's formula for the fermion determinant.  The generalization of
this procedure to $2K$ dimensions will be given below.

\newsec{$2K$ Dimensional Vertex Operators}

The generalization of what we have just said to four (and indeed to any
even
number of dimensions) is moderately straightforward.  We work in
Euclidean metric and place
a single Weyl fermion field
in a singular instanton field with delta
function topological charge, normalized
so that it creates a single unit of
chiral charge at the point $x$.  Such an instanton can be constructed
out of two dimensional instantons as follows.  Choose a set of $K$ mutually
orthogonal planes in $2K$ dimensions.  We will denote a generic point
in the space of all such ordered sequences of mutually orthogonal planes
by $P$, and planes belonging to the set by $i \in P$. Choose unit vectors
${\bf n}_i$ in each plane.
On each plane, place a two
dimensional instanton with delta function field strength on that plane.  If
$\omega_{\mu\nu}^i$, $i = 1 \ldots K$, are the area tensors of the planes
(i.e. constant antisymmetric tensors equal to the two dimensional
$\epsilon_{\mu\nu}$ when the indices are in the appropriate plane, and
zero
otherwise), and $$P_{\mu\nu}^i = \sum_{a=1}^2 \delta_{\mu a}\delta_{\nu
a}$$ (with $a$ the two orthogonal directions in the plane $i$) the
projection onto the plane, then the full gauge configuration is
\eqn\Ddinst{A_{\mu} = \sum_i A_{\mu}^i =
\sum_i {\omega_{\mu\nu}^i x^{\nu} \over {x^T P^i x}}}
This configuration has a unit topological charge located at $x = 0$.
Another way to write this configuration is to note that $A_{\mu}$ is
a singular pure gauge, $\partial_{\mu}\sum \theta_i (x)$, where the
$\theta_i$ are the angles between the projection of $x$ onto the plane
$i$ and the unit vector ${\bf n}_i$.

A vertex operator will be an expression of the form $e^{i \int
A_{\mu} J^{\mu}}$, where $J^{\mu}$ is the chiral fermion number current.
The fact that it depends on a choice of planes
shows that it will not be a scalar under rotation.  To understand how it
transforms we first note that there is in fact a whole multiplet
of vertex operators with the same topological charge even for a given
choice of planes.  Indeed we can change the sign of an even number of
the
planar instantons, without changing the topological charge.  The number
of different vertex operators with a given orientation of the planes
is thus $2^{K - 1}$.  which is precisely the number of components of a
Weyl spinor in $2K$ dimensions.  This motivates the following
construction:

The Weyl matrices in $2K$ dimensions are $2^{K-1} X 2^{K-1}$ dimensional
matrices combined together into the $2K$ vector $\sigma^{\mu} = (i,
\gamma^a)$ where $\gamma^a$ is a hermitian representation of the $2K -
1$ dimensional Dirac algebra.  They satisfy the algebra
\eqn\weylalg{\sigma^{\mu} {\sigma^{\dagger}}^{\nu} = \delta^{\mu\nu} + i
\Sigma^{\mu\nu}}
$\Sigma^{\mu\nu}/2$ are the generators of the Weyl representation of
$O(2K)$.  These matrices are hermitian and satisfy $(\Sigma^{\mu\nu})^2
= 1$ for each plane.  The Cartan subalgebra consists of the $\Sigma$
matrices for $K$ mutually orthogonal planes.  Note that one of these can ,
without
loss of generality be chosen to be one of the $(2K - 1)$ dimensional
Dirac matrices
while the others are commutators of independent  pairs of
the $2K - 2$ remaining
Dirac matrices.  Odd dimensional Dirac matrices satisfy the identity
\eqn\iden{\prod \gamma^a = 1}
so only $(K-1)$ of the eigenvalues of the Cartan generators can be chosen
independently.

 Now let $\varphi^n_{\alpha}$
be Weyl spinors which  are simultaneous eigenvectors of the Cartan
matrices.  Introduce a set of fermionic annihilation operators $C_n$.
 Define a vertex operator by
\eqn\Ddvertop{V_{\alpha} = (e^{i\int \sum\limits_i A_{\mu}^i \Sigma^i
J^{\mu}})_{\alpha}^{\beta} \sum_n c_n \varphi^n_{\beta}}
Here $\Sigma^i$ are the $\Sigma^{\mu\nu}$ corresponding to the
set of orthogonal planes we have chosen for the instanton.
As $\varphi$ runs over the independent eigenvectors of the Cartan
generators, these vertex operators run over the $2^{K - 1}$
possible choices of sign described above.

To discuss the transformation properties of the vertex operator, let us
imagine doing an $O(2K)$ transformation on the fundamental fermion
fields
in the functional integral.  The current transforms like a $2K$
dimensional
vector field, so by changing variables in the space time integral in the
exponent of the vertex operator, we see that the effect of this
operation is to rotate the vector potential.  A general rotation will
rotate the planes $P$ into another collection, $\Lambda P$,of
mutually orthogonal planes.
The vertex operator has become
\eqn\rotvertop{V_{\alpha}^{\prime} = (e^{i\int \sum\limits_{i_{R}}
 A_{\mu}^{i_R}\Sigma^i J^{\mu}})_{\alpha}^{\beta} \varphi_{\beta},}
where $i_R$ is the plane into which $i$ is rotated.
Now let ${\cal D}(R_{i_R})$ be the representation matrix in the Weyl
spinor representation of the rotation which takes the collection of
planes $[i]$ into the collection $[i_R]$.  Then
\eqn\rotvertopb{V_{\alpha}^{\prime} =
{\cal D}(R_{i_R})_{\alpha\gamma}(e^{i\int \sum\limits_i
A_{\mu}^{\Lambda P}\Sigma^{i_R} J^{\mu}})_{\gamma}^{\delta} {{\cal
D}^{\dagger}}(R_{i_R})_{\delta\beta}\varphi_{\beta}}
Now note that the matrix ${\cal D}^{\dagger}$, operating on the
eigenstates $\varphi$ of $\Sigma^i$, produces eigenstates of
$\Sigma^{i_R}$ with the same eigenvalue.
\eqn\eigentrans{{\cal D}^{\dagger}\varphi^n_{(P)} = e^{i \delta_n (\Lambda )}
\varphi^n_{(\Lambda P)}}

In order to cancel the phases in this tranformation we must allow the $c_n$
to transform under Euclidean rotations.  They are thus additional dynamical
variables in our system.
The operators $c_n$ are the analogs of the
zero mode fermion introduced in the sixth reference of \vert .
In two dimensions, bilinears in these zero mode fermions produce
the cocycle operators of the Frenkel-Kac construction.

With these rules, it is then clear that the vertex operator transforms
under an Euclidean rotation as a Weyl spinor:
\eqn\trans{V_{\alpha}(x) \rightarrow {\cal D}_{\alpha}^{\beta}(\Lambda )
V_{\beta}(\Lambda^{-1} x)}
\subsec{Green Functions}

In order to compute fermion correlation functions in terms of determinants in
back
ground multiinstanton fields, we use the same prescription that worked in
two dimensions.  If the gauge field is coupled to a Dirac fermion in a
vectorlike manner, there are two fermion zero modes for the Dirac operator in
a field with topological charge one,
and the Abelian analog of the 't Hooft interaction is bilinear in the
fermions.  The determinant in an instanton anti-instanton background
is thus proportional to $(x_I - x_{\bar{I}})^{-(4K - 2)}$, {\it i.e.} it
scales like the square of the fermion propagator\foot{Generally, this scaling
behavior is just the long distance limit of the determinant.  However, neither
the singular instantons nor the theory contain a scale, so apart from
renormalizations of the vertex operators, the long distance limit is the
whole function.}.  Note that this calculation is valid for any choice of
the eigenvalues of the $\Sigma_i$.  This means that if we define the
absolute value of the Weyl determinant to be the square root of the
Dirac determinant, then it is proportional to the unit matrix in Weyl
spinor space.  The same will not be true of the phase of the Weyl
determinant.  The full determinant will be a matrix
$D^{\alpha\gamma}_{\dot{\beta}\dot{\delta}} (x_I - x_{\bar{I}})$ and the
vertex operator two point
function will be
\eqn\twopt{<V_{\alpha}(x_I ) V_{\dot{\beta}}(x_{\bar{I}})> \propto
D^{\alpha\gamma}_{\dot{\beta}\dot{\delta}} (x_I - x_{\bar{I}}) <c_n \bar{c_m}>
\varphi_n^{\gamma}\bar{\varphi}_m^{\dot{\delta}}}
Here $\bar{c_m}$ are the zero mode operators for the antifermion field.

In order to complete the calculation, we must specify the two point function
of the zero mode operators. Define ${\bf N}\equiv {1\over\sqrt{K}}\sum
{\bf n}_i $.  We insist that
 \eqn\zeronorm{<c_n \bar{c_m}> \varphi_n^{\gamma}\bar{\varphi}_m^{\dot{\delta}}
= (N_{\mu}{\sigma^{\dagger}}^{\mu})_{\gamma\dot{\delta}}}
Note that in four dimensions any matrix connecting the spaces of the two spinor
representations is a linear combination of the $\sigma^{\dagger}$, so we are
only requiring that $\sqrt{K}{\bf N}$ have unit normalized
components in two orthogonal
planes.  In higher dimensions, we could in principle have obtained a more
complicated matrix.

Now let us return to the phase of the Weyl determinant.
The anomaly equation determines it to be
\eqn\weylphase{\ha\int [\sum (\theta_i (x - x_I )\Sigma^i + \theta_i
(x - x_{\bar{I}} )\bar{\Sigma^i} )\rho_{top}]}
(The matrices in this equation operate in the tensor product of the
two spinor representations.  Thus, the unbarred $\Sigma$ does not operate
on the barred indices and vice versa.).  The topological charge density
is
\eqn\topdensity{\rho_{top} = \delta^{2K} (x - x_I ) - \delta^{2K}
(x - x_{\bar{I}})}
The phase \weylphase thus contains some undefined pieces proportional
to $\theta_i (0)$ which we absorb into the normalization of the vertex
operators.  The finite part comes from the cross terms and has the form
\eqn\phasetwo{\ha\sum (\theta_i (x_I - x_{\bar{I}} )\Sigma^i + \theta_i
(x_I - x_{\bar{I}} )\bar{\Sigma^i} )}
This is just the Weyl representative of the rotation
 that rotates the vector ${\bf\Delta} ={\bf x_I - x_{\bar{I}}}$
into the direction of the vector ${\bf N}$.
Acting on $N_{\mu}{\sigma^{\dagger}}^{\mu}$ it produces ${\hat{\Delta}}_{\mu}
{\sigma^{\dagger}}^{\mu}$.  Combining this with the square root of the
Dirac determinant, we obtain precisely the free fermion propagator.

\newsec{Conformally Invariant Speculations}

In two dimensions, the vertex operator construction is the basis of a
remarkable
simplification of any conformal field theory with continuous global symmetries.
The ability to build operators of arbitrary transformation properties out of
the
currents themselves, suggests that the currents completely decouple from the
rest of the theory.  By multiplying an arbitrary charged operator by an
appropriate function of the currents one can construct a bleached operator
which
commutes with the currents.  The bleached operators are generally nonlocal,
 but
there are local functions of them which also commute with the currents.  These
form a separate conformal field theory with their own stress tensor.
To see this one constructs the affine-Sugawara stress tensor from the
currents, and
verifies that it is conserved, traceless and satisfies the Virasoro algebra.
The coset stress tensor obtained by subtracting $T_{Sugawara}$ from the
full stress tensor satisfies the Virasoro algebra with a complementary value
of the central charge.  It is the stress tensor of the bleached variables.
The Hilbert space of the theory is a tensor product of charged and bleached
sectors.

How much of this can we expect to generalize to higher dimensions?  One result
that certainly does not generalize is the formula for an abelian
current in terms
of a free boson field.  In higher dimensions, the currents of free fermions
have nonvanishing connected n point functions.  Indeed the
anomaly is just the connected 3 point function.
However, it is not inconceivable that the rest of the two dimensional
results generalize.  If we take any
subset of free fermion currents which forms a closed operator product algebra,
then the operator product expansion (OPE) of pairs of currents contains
a dimension four conserved symmetric traceless tensor $T_{\mu\nu}$ which is
a partial stress tensor and generates the correct equations of motion
for the currents.  Perhaps this is a special feature of the free fermion
theory but one may conjecture that it generalizes to interacting field
theories.

There is an obvious check of this conjecture which I have not yet
carried out.  Gauge theories with large numbers of colors and flavors
have perturbatively accessible fixed points.  This was first pointed out
in \ref\zaks{T.Banks, A.Zaks, {\it Nucl. Phys.}{\bf B196},
(1982),189.}\foot{Actually, in this
reference the perturbative
fixed point was accessed by taking the number of flavors
to be fractional.  This fractional flavor theory is well defined,
but not unitary.  I believe that it was D.Gross who first pointed out
that for large N one could achieve a perturbative fixed point with integer
number of colors and flavors, thus preserving unitarity.}
.  One can check the OPE of two currents at these perturbative but nontrivial
fixed points, to see whether the dimension of the symmetric traceless
tensor operator in this OPE changes.  If it remains equal to four, then
the stress tensor at the nontrivial fixed point will decompose into
two commuting pieces.

In this case we would conjecture that a four dimensional conformal field
theory
containing fermion fields would decompose into a tensor product of a \lq\lq
conformal current algebra'' theory and a bleached theory.  Note the restriction
to theories containing fermions.  A crucial part of the argument is our
ability to build charged fields out of neutral ones.  As we have seen, this
is a consequence of the anomalous violation of fermion current conservation
in the presence of instantons.  I know of no analogous phenomenon for
purely bosonic canonical variables in four dimensions.

\newsec{Applications?}

The higher dimensional bosonization formulae that we have constructed
are certainly more intricate than their two dimensional counterparts.
How do they compare in utility?  Certainly the formula hints at
a more fundamental understanding of what fermions really are, but
it remains to be seen whether it can lead to solutions of models
that cannot be understood in any other manner.

The most promising avenue of research in this regard, and one of the primary
motivations for the present work, is the profound generalization of
electric magnetic duality discovered by Seiberg in supersymmetric nonabelian
gauge theories\ref\nati{N.Seiberg,{\it Nucl. Phys.}{\bf B435},(1995),129.}.
At the core of Seiberg's argument for duality
is a conformal field theory that appears to be infinitely strongly coupled
in terms of the original (electric) variables.

Our bosonization formulae can be generalized to multifermion theories that
contain nonabelian global symmetries which we can gauge.  As in two dimensions,
the theory at finite gauge coupling looks ugly when written in terms of
bosonized variables (the cartan subalgebra of gauge currents)\foot{We have
constructed the higher dimensional analog of abelian but not of nonabelian
bosonization.}, and is not conformally invariant.  However, the
infinitely strongly coupled theory is formally obtained by modding out
the free theory by the gauge group.  That is, locally gauge invariant
operators which do not contain the gauge fields retain their free field
form, and gauge invariance is simply the statement that the nongauge invariant
operators do not act on the physical Hilbert space.

By analogy with orbifold theories in two dimensions, one may then expect to
find
that there exist operators in the free theory which are not local with respect
to the fundamental fermion fields, but are local with respect to all
gauge invariant functions of them.  I conjecture that these may be
 the ``magnetic quarks''
of Seiberg's duality transformation.

I have not yet been able to construct such magnetic operators.  Furthermore,
it is clear that there is more to the story than the outline above.  Seiberg's
construction leads us to expect free magnetic gauge bosons at the conformal
point.  How are these to be constructed?  In addition, his dual theory
contains elementary fields corresponding to certain gauge invariant fermion
bilinears.  These have different dimension at the free magnetic fixed point
than they do at the free electric fixed point, contradicting the orbifold
analogy propounded above.  It is perhaps superfluous to state that more
research along these lines is necessary .
\vfill\eject
\centerline{\bf Acknowledgements}
Much of this paper was written when the author was a visitor at the Weizmann
Institute of Science during January of 1995. It was completed at the Stanford
Linear Accelerator Center in March of 1995. I would like to thank the
members of the Department of Particle Physics of the Weizmann Institute, and
of the theory groups at SLAC, Stanford, and the University of California
at Santa Cruz for their hospitality.  I would
also like to thank D.Friedan for a conversation about fermion zero modes,
and M.Peskin for reminding me how to compute the Dirac determinant in an
instanton-antiinstanton background.
\listrefs
\end